\documentclass[11pt,prd,tightenlines,nofootinbib,notitlepage,preprintnumbers]{revtex4-1}
\usepackage{hyperref}

\usepackage{psfrag}
\usepackage{graphicx}
\usepackage{amsmath}
\usepackage{amssymb}
\usepackage{amsthm}
\usepackage{psfrag}
\usepackage{graphicx}

\begin{document}
\preprint{LMU-ASC 71/10}

\def\be{\begin{equation}}
\def\ee{\end{equation}}
\def\bea{\begin{eqnarray}}
\def\eea{\end{eqnarray}}
\def\ba{\begin{array}}
\def\ea{\end{array}}
\def\bse{\begin{subequations}}
\def\ese{\end{subequations}}

\def\pd{{\partial}}

\def\a{{\alpha}}
\def\b{{\beta}}
\def\g{{\gamma}}
\def\m{{\mu}}
\def\n{{\nu}}
\def\d{{\delta}}
\def\e{{\epsilon}}

\def\L{{\mathcal{L}}}
\def\E{{\mathcal{E}}}

\title{Galileon accretion}
\author{E. Babichev}
\affiliation{Arnold-Sommerfeld-Center for Theoretical Physics,
Department f\"ur Physik, Ludwig-Maximilians-Universit\"at M\"unchen,
Theresienstr. 37, D-80333, Munich, Germany}

\begin{abstract}
We study steady-state spherically symmetric accretion of a galileon field onto a Sch\-warz\-schild black hole in the test fluid approximation.
The galileon is assumed to undergo a stage of cosmological evolution, thus setting a non-trivial boundary condition at spatial infinity.
The critical flow is found for some parameters of the theory. 
There is a range of parameters when the critical flow exists, but the solution is unstable.
It is also shown that for a certain range of parameters the critical flow solution does not exist.
Depending on the model the sound horizon of the flow can be either outside or inside of the Schwarzschild horizon. 
The latter property may make it problematic to embed the galileon theory in the standard black hole thermodynamics. 
\end{abstract}

\maketitle



\section{Introduction}
Galileon is a scalar field theory, introduced in \cite{Nicolis:2008in}. This theory has a remarkable property: 
its Lagrangian contains higher-order derivative terms, while the equations of motion are of only second derivatives of the field.
The galileon theory is invariant under the galilean transformation of the field. 
It was originally introduced as a generalization of the scalar field theory, 
left over in the decoupling limit \cite{Nicolis:2004qq} of Dvali-Gabadadze-Porrati (DGP) model \cite{Dvali:2000hr} and which  
allows the implementation of the Vainshtein mechanism \cite{Vainshtein:1972sx}.
The covariant version of galileon (covariant galileon) was found in \cite{Deffayet:2009wt}, 
where a model yielding only up to second-order derivatives in both field and metric was constructed.
Other aspects of the galileon and related ideas were developed later
\cite{Deffayet:2009mn,*deRham:2010eu,*Deffayet:2010zh,*Padilla:2010de,*Hinterbichler:2010uq,*Padilla:2010tj,*Andrews:2010kx,*Goon:2010fk}.

The galileon and related models were also applied for construction of cosmological models \cite{Chow:2009fm}
(a more generic scalar-tensor model was suggested in \cite{Babichev:2009ee} where the gravity is modified in infra-red, 
while the nonlinear kinetic coupling provides the restoration of general relativity).
More recently, several galileon-related models were studied in the cosmological context, see e.g.
\cite{Deffayet:2010qz,
Creminelli:2010ba,*Felice:2010fk,*Kobayashi:2010cm,*DeFelice:2010nf,
Silva:2009km,*Kobayashi:2009wr,*Kobayashi:2010wa,*Gannouji:2010au,*Ali:2010gr}.

The galileon model is interesting in several aspects. 
First of all, this is a theory with non-quadratic kinetic coupling.
This leads to propagation of perturbations in an effective metric, which is generically different from the gravitational one.
In particular, the propagations may have sub- or superluminal velocities on non-trivial backgrounds.
And indeed, as it was shown in \cite{Nicolis:2008in}, galileon generically contains superluminal excitations 
(the same happens in for a scalar left over in the decoupling limit of DGP \cite{Adams:2006sv}).
The existence of faster-than-light propagation on non-trivial backgrounds does not necessary imply inconsistencies  
\cite{Babichev:2007dw,Bruneton:2006gf,*Bruneton:2007si,*Geroch:2010fk}.
In particular, ``superluminal'' DBI inflation \cite{Mukhanov:2005bu} or accreting k-essence \cite{Babichev:2006vx,Babichev:2007wg}
are free of any time paradoxes
(Note, that in \cite{Adams:2006sv}, however, it was argued that superluminality of certain theories might indicate the problem for embedding 
them into a UV complete theory).
Another interesting feature of galileon was recently discussed in \cite{Deffayet:2010qz,Creminelli:2010ba,*Felice:2010fk,*Kobayashi:2010cm,*DeFelice:2010nf}: 
cosmological models with the phantom behavior were constructed, and it was argued that 
perturbations are ghost-free for some choice of parameters and initial conditions.

In the original non-covariant version of the galileon it was found that the coefficient of the Lagrangian must be constrained, 
depending on the background \cite{Nicolis:2008in}. The check was performed by constructing static spherically symmetric solutions 
with a source in the de Sitter universe. 
It is thus seems important to continue this route and to test the galileon model further against pathologies. 
In particular, the covariant version of the galileon must be checked for existence of solutions and their stability.
One of the useful area to test theories for unusual/pathological behavior is black hole physics. 
Indeed, previously the study of models in the black hole background has revealed interesting (and sometimes undesired) features of 
underlying models, e.g. it was shown that 
the phantom field accretion decreases the black hole mass \cite{Babichev:2004yx,*Babichev:2005py,*Babichev:2008jb};
the ghost condensate plus a black hole forms a perpetuum mobile of the 2nd kind \cite{Dubovsky:2006vk,*Eling:2007qd} 
(however  in \cite{Mukohyama:2009rk} it was argued that ghost condensate does not violate the second law of thermodynamics);
and the superluminal k-essence allows one to send signals from inside of the event horizon of a black hole \cite{Babichev:2006vx,*Babichev:2007wg}.

In this paper we consider the galileon accretion onto a Schwarzschild black hole. 
The paper is organized as follows.
In Sec.~2 we specify the action of the theory,
give equations of motion and the energy-momentum tensor for different types of galileon. 
Section~3 is devoted to the galileon of the certain type, namely, containing only the terms 
left over in the decoupling limit of DGP.
We construct the family of solutions describing a steady-state inflow of the scalar field.
The unique physical solution is fixed by the requirement that it is regular both at the Schwarzschild and sound horizons.
In Sec.~4 we widen our study to include  another form of the galileon term.
In Sec.~5 we briefly summarize our results, discuss applicability of our study and physical consequences.
In particular, for the galileon of the DGP type we find no constrains on the Lagrangian. 
When another galileon term is included, we obtain an additional constraint as compared to \cite{Nicolis:2008in}.
For a certain choice of parameters, the galileon  forms the sound horizon inside the Schwarzschild one, 
thus giving an opportunity to send signals from the inside of the Schwarzschild horizon.
We also discuss  accretion of the ``phantom'' type of galileon.

\section{Model}
The general scalar field covariant galileon action reads \cite{Deffayet:2009wt}
\be
\label{action0}
S_\pi = \int d^4 x\sqrt{-g}\,\L_\pi,
\ee
where Lagrangian density for the
 can be written as a linear combination of terms,
\be
\label{L}
\L_\pi = \sum_{i=1}^{i=5} c_i\L_i,
\ee
where, 
\be
\label{L1234}
\begin{aligned}
\L_1 &= \pi, \;\;
\L_2 = \pi_{;\mu} \pi^{;\mu}, \;\;
\L_3 =  \pi_{;\mu} \pi^{;\mu} \Box\pi, \\
\L_4 &=  \left(\pi_{;\a}\pi^{;\a}\right)\left[ 2\left(\Box\pi\right)^2 - 2 \left(\pi_{;\mu\nu}\pi^{;\mu\nu} \right) -\frac12 \pi_{;\mu}\pi^{;\mu} R \right],
\end{aligned}
\ee
and the term $\L_5$ has a more complicated structure and contains higher order derivatives in $\pi$, we are not going to study 
it in this paper\footnote{Throughout this paper we use the signature $(-,+,+,+)$. 
We also follow here the notations of Ref.~\cite{Deffayet:2009wt}, which are different to ones in \cite{Nicolis:2008in}. 
Namely, neglecting gravity, and up to the integration by parts, 
$\L_2 = -2 \L_2^{NRT}$, $\L_3 = -2 \L_3^{NRT}$ and $\L_4 = -4 \L_4^{NRT}$, where $\L_i^{NRT}$ are the definitions of \cite{Nicolis:2008in}.
To avoid confusions with signs we also assume that our full action, including gravity, has the overall minus sign compared to \cite{Nicolis:2008in},
so that, e.g. the constraint $d_2>0$ in \cite{Nicolis:2008in}, for asymptotically flat space-time implies $c_2>0$ in our notations.}. 
We will also set $c_1$ to zero in the following study, i.e. the ``potential'' term is not included.
Otherwise, if $c_1\neq 0$, there is no steady-state solution for the accretion (this is similar to the case of the k-esence field see, e.g. \cite{Akhoury:2008nn}).
The constant $c_2$ in (\ref{L}) is dimensionless, $c_3$ has dimension $-3$, in the context of the DGP
model this coefficient is usually written as $r_c/M_{\rm Pl}^2$ where $r_c$ is the so called cross-over scale and 
$M_{\rm Pl}$ is the Planck mass. 
The coefficient $c_4$ has mass dimension $-6$.

The energy-momentum tensor derived from (\ref{L}), (\ref{L1234}) as $T^{(i)}_{\m\n} \equiv 2/\sqrt{-g} \left( \delta S_{(i)}/\delta g^{\m\n}\right)$ is
\bse
\label{emt}
\begin{align}
T_{\m\n}^{(2)}&=  2\pi_{,\m} \pi_{,\n} - g_{\m\n}\left(\pd\pi\right)^2, \label{emt2}\\
T_{\m\n}^{(3)}&=  2\pi_{,\mu} \pi_{,\nu}\Box\pi - 2\pi_{,(\mu}\nabla_{\n)} \left(\pd\pi\right)^2
+ g_{\m\n}\pi^{,\a} \nabla_\a \left(\pd\pi\right)^2, \\
T_{\mu \nu}^{(4)} &=
-8
\left(\Box \pi\right)\pi^{;\rho}\bigl[\pi_{;\mu}\,\pi_{;\rho\nu}+\pi_{;\nu}\,\pi_{;\rho\mu}\bigr]
	+4  \left(\Box \pi\right)^2 \left(\pi_{;\mu}\,\pi_{;\nu}\right)
	-4  \left(\Box \pi\right)\left(\pi_{;\lambda}\,\pi^{;\lambda}\right) \left(\pi_{;\mu\nu}\right)
\nonumber \\
& -8
	\left(\pi_{;\lambda}\,\pi^{;\lambda\rho}\,\pi_{;\rho}\right) \left(\pi_{;\mu\nu}\right)
	+8 \left(\pi^{;\lambda}\,\pi_{;\lambda\mu}\right)\left(\pi^{;\rho}\,\pi_{;\rho\nu}\right)
	-4  \left(\pi_{;\lambda\rho}\,\pi^{;\lambda\rho}\right)
	\left(\pi_{;\mu}\,\pi_{;\nu}\right)
\nonumber \\
&+4 
	\left(\pi_{;\lambda}\,\pi^{;\lambda}\right)
	\left(\pi_{;\mu\rho} \, \pi^{;\rho}_{\hphantom{;\nu}\nu}\right) + 8 \,\pi_{;\lambda}\,\pi^{;\lambda\rho} \bigl[\pi_{;\rho\mu}\,\pi_{;\nu}
	+\pi_{;\rho\nu}\,\pi_{;\mu}\bigr] +2 \left(\Box \pi\right)^2 \left(\pi_{;\lambda}\,\pi^{;\lambda}\right)g_{\mu \nu}
\nonumber \\
&+ 8  \left(\Box \pi\right)
	\left(\pi_{;\lambda}\,\pi^{;\lambda\rho}\,\pi_{;\rho}\right) g_{\mu \nu}
	-8 \left(\pi_{;\lambda}\,\pi^{;\lambda\rho}\,\pi_{;\rho\sigma}\,\pi^{;\sigma}\right) g_{\mu \nu}
\nonumber \\
&-2 \left(\pi_{;\lambda}\,\pi^{;\lambda}\right)
	\left(\pi_{;\rho\sigma}\,\pi^{;\rho\sigma}\right) g_{\mu \nu} -2  \left(\pi_{;\lambda}\,\pi^{;\lambda}\right)
	\left(\pi_{;\mu}\,\pi_{;\nu}\right) R + \frac{1}{2} \left(\pi_{;\lambda}\,\pi^{;\lambda}\right) \left(\pi_{;\rho}\,\pi^{;\rho}\right)R g_{\mu \nu} 
\nonumber \\
& +4 \left(\pi_{;\lambda}\,\pi^{;\lambda}\right) \pi^{;\rho} \bigl[R_{\rho\mu}\,\pi_{;\nu}
	+R_{\rho\nu}\,\pi_{;\mu}\bigr] - \left(\pi_{;\lambda}\,\pi^{;\lambda}\right) \left(\pi_{;\rho}\,\pi^{;\rho}\right) R_{\mu \nu}
\nonumber \\
& - 4 \left(\pi_{;\lambda}\,\pi^{;\lambda}\right)
	\left(\pi_{;\rho}\,R^{\rho \sigma}\,\pi_{;\sigma}\right) g_{\mu \nu}
	+4 \left(\pi_{;\lambda}\,\pi^{;\lambda}\right) \left(\pi^{;\rho}\,\pi^{;\sigma}\,R_{\mu\rho\nu\sigma}\right). \label{emt4}
\end{align}
\ese
Note that the expression for ${ T}^{(4)}_{\mu \nu}$ differs from the one derived in \cite{Deffayet:2009wt} 
by factor $-2$, because of the difference in the definition for the energy-momentum 
tensor\footnote{In \cite{Deffayet:2009wt} the energy-momentum tensor was defined as $T^{\m\n} \equiv 1/\sqrt{-g} \left( \delta S/\delta g_{\m\n}\right)$.}.
Equations of motion are obtained by the variation of (\ref{action0}) with respect to $\pi$, $\E_{(i)} \equiv (-g)^{-1/2}\delta S_{(i)}/\delta\pi =0$,
with
\be
\label{eomDGP}
\begin{aligned}
\E_2 &=  - 2 \Box\pi, \nonumber \\
\E_3 &=  -2 \left( \left(\Box\pi\right)^2 - \left(\nabla\nabla\pi\right)^2 -  R^{\m\n}\pi{,_\m}\pi_{,\n} \right), \nonumber\\
\E_4 &= 
 - 4 \left(\Box \pi \right)^3
-8 \left(\pi_{;\mu}^{\hphantom{;\mu}\nu}\,\pi_{;\nu}^{\hphantom{;\nu}\rho}\,\pi_{;\rho}^{\hphantom{;\rho}\mu}\right)
+12 \left(\Box \pi\right) \left(\pi_{;\mu\nu}\pi^{;\mu\nu} \right)
+ 2 \left(\Box \pi\right) \left(\pi_{;\mu}\,\pi^{;\mu}\right)  R\nonumber\\
&+ 4 \left(\pi_{;\mu}\,\pi^{;\mu\nu}\,\pi_{;\nu}\right) R
+8 \left(\Box \pi\right) \left(\pi_{;\mu} \,R^{\mu \nu}\,\pi_{;\nu}\right)
- 4\left(\pi_{;\lambda}\,\pi^{;\lambda}\right)
\left(\pi_{;\mu\nu}\,R^{\mu \nu}\right)\nonumber \\
&   -16 \left(\pi_{;\mu}\,\pi^{;\mu\nu}\,R_{\nu\rho}\,\pi^{;\rho}\right)
- 8 \left(\pi_{;\mu}\,\pi_{;\nu}\,\pi_{;\rho\sigma}\,R^{\mu \rho \nu \sigma}\right). \nonumber
\end{aligned}
\ee
where $R_{\m\n}$ and $R_{\m\n\alpha\beta}$ are the Ricci and Riemann tensors correspondingly.

In what follows it will be useful to introduce dimensionless quantities in order to simplify formulae. 
Having in mind that we will apply our formalism to the problem of accretion, 
a convenient rescaling is
\be
\label{rescale}
x^\mu\to r_g x^\mu,\quad  \quad \pi \to Cr_g \pi,
\ee
where $r_g=2M$ is the gravitational radius of a black hole, and the constant $C$ (with mass dimension 2)
will be associated with the cosmological value of  $\pd_t\pi$, see below.

\section{Case $\L_2 + \L_3$}

In this section we consider the accretion of a galileon, having nonzero $\L_2$ and $\L_3$, while other terms set to zero. 
This type of action (up to coefficients in front of $\L_2$ and $\L_3$) appears as the effective scalar field action, 
left over after taking the decoupling limit in DGP model of gravity.
In the rescaled units the action reads,
\be
\label{L2L3action}
S_\pi = r_g^4 C^2\int d^4 x\sqrt{-g}\,\left[ \e \left(\pd\pi\right)^2 + \kappa \left(\pd\pi\right)^2\Box\pi \right],
\ee
where $\e = 0,\, \pm1$, $\kappa = C c_3/r_g$ and we let $\kappa$ to be positive or negative.
Positive $\e$ corresponds to the canonical kinetic term, while positive both $\e$ and $\kappa$ 
correspond to the decoupling limit of DGP\footnote{Note that the overall sign in the action is opposite to \cite{Nicolis:2008in},
so that in our notations the canonical kinetic (no-ghost) term is $+(\pd \pi)^2$.}. 
Note that the the constraints from \cite{Nicolis:2008in} imply $\e>0$ and $\kappa\geq 0$ in the asymptotically flat space-time.
In this case, for the static solution far from the source, $\pi' \propto r^{-2}$ (such that $\pi'>0$), and the perturbations $\delta\pi$
on the background $\pi$ propagate superluminally. In this paper, we, however, do not restrict ourselves to only positive $\e$ and $\kappa$.

The equations of motion derived from (\ref{L2L3action}) are
\be
\label{eom}
\nabla_\mu j^\mu = 0,\quad j_{\mu} \equiv 2 \e\, \pi_{,\m} + \kappa\, \left(2 \pi_{,\m}\Box\pi - \pd_\mu \left(\pd\pi\right)^2\right),
\ee
or, 
\be
\label{eom1}
\e\,\Box\pi +\kappa\left( \left(\Box\pi\right)^2 - \left(\nabla\nabla\pi\right)^2 -  R^{\m\n}\pi{,_\m}\pi_{,\n} \right)= 0.
\ee

We will also need the equation for perturbations $\delta\pi$ on a non-trivial background $\pi(t,x)$, 
in the limit of high frequencies\footnote{Not to be confused with ``perturbation'' which sometimes used in a different context:
as a deviation of the solution from the homogeneous configuration due do the presence of a static source.}.
From (\ref{eom1}), we find the equation,
\be
\label{eomG}
G^{\mu\nu}\nabla_\mu\nabla_\nu\d\pi= 0,
\ee
where
\be
\label{G}
G^{\m\n} = \left(\e+2\kappa\Box\pi\right)g^{\m\n} - 2\kappa \nabla^\m\nabla^\n\pi.
\ee
Note that in deriving (\ref{eomG}), (\ref{G}) we neglected the term $\sim \pd\pi\pd\pi \delta R$,
although there is a contribution of this term in the effective metric (\ref{G}), see \cite{Deffayet:2010qz}. 
However, this term is suppressed by  $M_{\rm Pl}^2$ \cite{Deffayet:2010qz}, 
so we can safely neglect it in our calculations (see also the discussion in Sec.~5).

The propagation vectors for small perturbations can be found from the relation (see, e.g. \cite{Babichev:2007dw}),
\be
\label{vectors}
\tilde{G}_{\mu\nu}\eta^\mu\eta^\nu = 0,
\ee
where $\tilde{G}_{\mu\nu}$ is the inverse matrix to $G^{\mu\nu}$, $\tilde{G}_{\mu\nu} G^{\mu\nu} =1$.

Since we are interested in solutions of a scalar field including some region inside the Schwarzschild horizon, 
we will be needing a regular (on the Schwarzschild horizon) coordinate system.
E.g. we can take ingoing Eddington-Finkelstein (EF) coordinates.
The EF coordinates $(v,\, r)$ are connected to the usual Schwarzschild coordinates $(t,\, r)$ in the following way,
\be
\label{EFcoord}
v= t+ \int\frac{dr}{f},\quad r=r,\nonumber
\ee
where 
\be
\label{f}
f=1-\frac{1}{r}, 
\ee
in the rescaled units.
The metric of a Schwarzschild black hole metric in EF coordinates reads,
\be
\label{EF}
d s^2 =  -f d v^2 + 2d vdr + r^2 d \Omega,
\ee
To describe a steady-state accretion we take the ansatz,
\be
\label{ansatz}
\pi(v,r) = v - \int\frac{dr}{f} + \psi(r).
\ee
Note that by taking the ansatz (\ref{ansatz}) we used the freedom in choosing the rescaling in (\ref{rescale}),
thus we took $C$ to be equal $\pd_t\pi$ at the spatial infinity,
\be
\label{C}
C = \pd_t\pi |_{r=\infty} = \pd_v\pi|_{r=\infty},\nonumber
\ee
thus setting the coefficient in front of $v$ in (\ref{ansatz}) to unity.
The term $\int f^{-1}dr$ in (\ref{ansatz}) was introduced for convenience, such that for the homogeneous solution (with no black hole in the Universe),
$\psi(r) = {\rm const}$.
Since the current depends only on $r$, Eq.~(\ref{eom}) can be integrated once to give
\be
\label{1int}
r^2 j^r = A,
\ee
where $A$ is a constant determining the total flux (to be fixed below).
For the ansatz (\ref{ansatz}) the $r$-components of the current reads,
\be
\label{jr}
j^r =  2\e\, f \psi '+ \kappa  \left(-\frac{f'}{f}+f f' \psi '^2+\frac{4 f^2 \psi '^2}{r}\right).
\ee
Eqs.~(\ref{1int}), (\ref{jr}) can be obtained in a different way, namely, by integrating the 
equation for the energy-momentum conservation, $T^{\mu\nu}_{\hphantom{\mu;};\nu}=0$, which gives $r^2T_v^r = {\rm const}$,
where for the ansatz (\ref{ansatz}) one can find $T_v^r = j^r$. 
The Eqs.(\ref{1int}) and (\ref{jr}) form an algebraic equation on $\psi'$. 
The solution contains constant $A$ as a free parameter,
$
\psi' = \psi'(A,r).
$
The physical solution will be chosen by the requirement that it is neither singular at the Schwarzschild horizon nor at the sound horizon. 
This seemingly ``weak''  requirement choses the solution uniquely, which in hydrodynamics is called transonic branch. 
To fix the transonic solution one needs to find a position of the sound horizon,
i.e. one needs to study  perturbations on the background solution $\pi_A(v,r)$.
There are generically two solutions of (\ref{1int}) and (\ref{jr}), 
\be
\label{sol12}
\psi'_{(2,3)}=
 - \frac{ \e\,r^2 f\pm
	\sqrt{\e^2\, r^4 f^2 +\kappa r \left(A f +\kappa r^2 f' \right)\left(r f'+4 f\right) }}
		{ \kappa r f \left(r f'+4 f\right)},
\ee
where the subscript $(2,3)$ means that the solution was obtained for the theory with $\L_2$ and $\L_3$ terms in the Lagrangian.
Below we consider solutions different cases corresponding to the choice of coefficients in the action (\ref{L2L3action}).
\subsection{Canonical kinetic term, $\e= 1$, $\kappa = 0$.}
The standard case is recovered from (\ref{sol12}) by setting $\e=1$ and taking the limit  $\kappa\to 0$ for the  solution with the minus sign
(alternatively, one can solve (\ref{1int}) and (\ref{jr}) setting $\kappa=0$), 
\be
\label{solst}
\psi'_{\rm can} (A) = \frac{A}{2r^2 f}.
\ee
It is not hard to see that the physical solution corresponds to the choice $A=2$ in (\ref{solst}).
The sound horizon $r_*$ coincides with the Schwarzschild horizon, $r_*=1$,
which is not surprising, since the speed of excitations for the the canonical term is the speed of light.
The solution for $\pi'(r)$ is shown in Fig~\ref{fig DGP}.
%
\subsection{No canonical  term, $\e = 0$, $\kappa \neq 0$.}
%
It is interesting to look at the case when the quadratic kinetic term is absent. This case is simpler than the generic case (which we consider below), 
and in meantime it gives an insight to the full problem. From Eq.~(\ref{sol12}), setting $\e=0$ and substituting $f(r)$ from (\ref{f}), we obtain,
\be
\label{solNC}
\psi'_{(3)}(A)=
 \mp 
	\frac{1}{r-1}\sqrt{\frac{r\left[A(r-1) + \kappa r\right]}{\kappa(4r-3)}}.
\ee
For the solution with the plus sign 
the singularity coming from the denominator of (\ref{solNC}) at $r=1$
is cancelled by the term $\left(-\int f^{-1}dr\right)$ in (\ref{ansatz}) for any $A$; 
while the solution with the minus sign leads to a divergency of  physical quantities. 
Another potentially dangerous point is at $r=3/4$.
The only way to avoid the singular behavior at this point is to to choose $A=3\kappa$. 
Thus,
\be
\label{solNCph}
\psi'_{{\rm (no\, can)}}=\frac{\sqrt{r}}{r-1}
\ee
is the physical solution, in Fig.~\ref{fig DGP} the corresponding solution for $\pi'$ is shown.
Let us now analyze propagation of perturbations. From (\ref{G}) setting $\e=0$ one can find,
\be
\label{GNC}
G^{00}=-2 \kappa  \left(\frac{f'}{f^2}+\psi''\right), \; 
G^{01} = \kappa  \left(f' \psi'-\frac{f'}{f}+\frac{4 f\psi'}{r}\right),\;
G^{11}= \frac{\kappa  f \left(rf'+4 f\right) \psi'}{r}.
\ee
Then solving the equation (\ref{vectors}) with the covariant metric $\tilde{G}_{\m\n}$, being the inverse of (\ref{GNC}), 
we find two characteristics of the differential equation, $\eta\equiv dv/dr$,
\be
\label{etaNC}
\eta_{(3)} = \frac{\left(2\pm\sqrt{3}\right)r}{\left(1+\sqrt{r}\right)\left(2\sqrt{r}\mp\sqrt{3}\right)}.\nonumber
\ee
The divergence of one the characteristic $\eta_{(3)}$ (with the minus in the denominator) at $r=3/4$, 
indicates the position of the sound horizon, $r_* = 3/4$. Note that in this case the sound horizon is inside the Schwarzschild one.
The other propagation vector, which corresponds to the perturbations being sent into the black hole, is finite everywhere at $0<r<\infty$.

\subsection{Generic case, $\e \neq 0$, $\kappa \neq 0$.}

Let us now turn to the generic case, i.e. when both terms in (\ref{L2L3action}) are nonzero.
First we consider the quadratic term in action (\ref{L2L3action}) to be canonical, $\e=1$.
The solution with the plus sign (\ref{sol12}), for any $A$, at infinity  is approximated by
\be
\label{sol1inf}
\psi'_{(2,3)+} = -\frac{1}{2\kappa} r + {\cal O}(1),\nonumber
\ee
thus violating our assumption on the homogeneity of the solution at the spatial infinity. 
Meantime it is regular at $r=3/4$.
The other solution, on the contrary, is well-behaved at infinity,
\be
\label{sol2inf}
\psi'_{(2,3)-} = \frac{1}{2r^2}\left(A+\kappa\right) + \frac{1}{r^3}\left(\frac{A}{2} + \kappa\right) + \mathcal{O}\left(r^{-4}\right).
\ee
However, the solution (\ref{sol2inf}) diverges at $r=3/4$ along with the characteristics $\eta_{(2,3)}$ for the generic 
choice of $A$. Thus, similar to the standard problem of fluid accretion,
one needs to find such a value of $A$ that at some point $r=r_*$ the solutions $\psi_{(2,3)\pm}$ and their derivatives match. 
The surface $r=r_*$ (called the critical point), 
is analogous to the acoustic horizon, from inside of which perturbations of $\pi$ cannot  escape to the asymptotically flat regions (see below). 
At the critical point  one of the propagation vectors $\eta_{(2,3)}$ diverges, i.e. in our case $\tilde{G}_{vv}=0$, 
implying $G^{rr}=0$. 
However, up to the non-relevant conformal factor, 
$G^{rr}$ is also being the coefficient in front of $\psi''$ in the equation of motion for the background solution. 
In order for the solution to be smooth at the sound horizon, the rest of equation of motion for $\psi$ 
(containing $\psi$, $\psi'$ and terms) should also be zero at the sound horizon.
Thus the procedure is as follows. 
Differentiating (\ref{1int}) once, we obtain the second derivative equation on $\psi$. Then one can express $\psi''$ in terms 
of $\psi'$ (and there is no $A$ in the expression, because it disappears during the differentiation). 
Now, the critical point is where both denominator and numerator are equal to zero. 
Solving simultaneously these two equations, one obtains an algebraical equation $P(r)=0$, where $P(r)$ is a polynomial 6th order in our case.
The position of the sound horizon, $r_*$, can be then found numerically from this equation. 
After finding $r_*$, the constant of integration $A$ is fixed.
There are generically two possibilities, one of them corresponds to (physical) inflow and the other to (unphysical) outflow.
\begin{figure}[t]
\begin{center}
\includegraphics[width=0.5\textwidth]{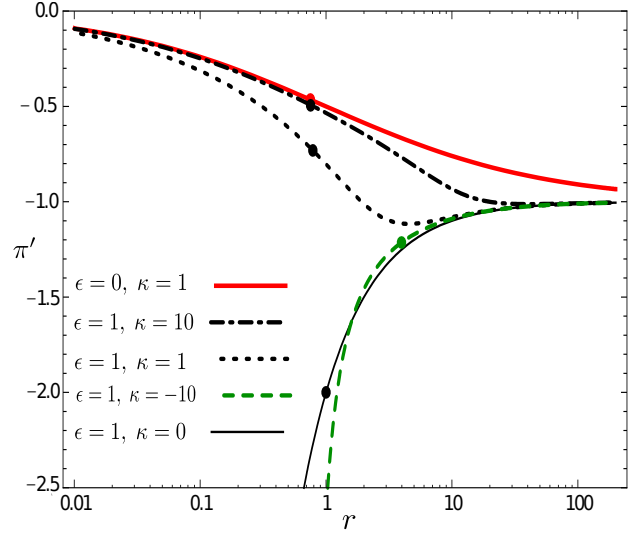}
\caption{In the case $\L\sim \L_2 +  \L_3$, solutions for $\pi' = -1/f+\psi'(r)$ are shown for different parameters of the model. 
The positions of sound horizon are shown by the dots. 
In case $\kappa <0$ the sound horizon is outside the Schwarzschild horizon, while for $\kappa>0$ the sound
horizon is inside the Schwarzschild horizon.}
\label{fig DGP}
\end{center}
\end{figure}

For a positive $\kappa$, corresponding to the decoupling limit of the DGP model,
the sound horizon is inside the Schwarzschild horizon.
The latter could be anticipated, since in this case
the perturbations of the scalar field on nontrivial background of $\pi$ are superluminal. 
There is a regular solution in the whole range of $r$, $0<r<\infty$, see Fig.~\ref{fig DGP}.
For $\kappa<0$, on the contrary, the sound horizon is outside the Schwarzschild horizon. 
The solution is smooth outside of the Schwarzschild horizon, and it becomes singular at some point $r<3/4$, 
which can be seen by analyzing the square root in (\ref{sol12}).
This solution is nevertheless physical, since the pathological behavior happens inside of both acoustic and Schwarzschild 
horizons.

After fixing the critical inflow solution, 
one can analyze the characteristics of the equations for perturbations, 
$\eta_{(2,3)}$  and to check that $r_*$  is indeed the sound horizon.
Fig.~\ref{fig DGP} shows behavior of the solutions for $\pi'$ for various cases.

Note that one can also relax the requirement of positivity of $\e$ (being the requirement that the Lagrangian is having healthy canonical term)
and to consider the case when the canonical kinetic term has a negative sign (ghost-like).
Such a Lagrangian was, e.g. considered in the cosmological context in \cite{Deffayet:2010qz}.
In our study, this situation can be easily reduced to one of the studied cases. 
Indeed, if both $c_2$ and $c_3$ are negative, then action is equivalent to the case
$c_2 > 1$, $\kappa >0$, up to the total sign in front of the action, so that the equations of motion do not change. 
The action with $c_2<0$, $c_3<0$, is equivalent to the studied above case $c_2 > 0$, $c_3 <0$, up to the total sign in front of the action.
Note that we do not address the problem of ghosts here, we are only interested in the existence and stability of the classical solutions.

\section{Case of non-zero $\L_4$}
\label{L4}
%
In this section we consider the case of nonzero $\L_4$,
\be
\label{L2L4}
S_\pi = r_g^4 C^2\int d^4 x\sqrt{-g}\,\left[ \e \left(\pd\pi\right)^2 + \bar{\kappa}  
\left(\pd\pi\right)^2\left( 2\left(\Box\pi\right)^2 - 2 \left(\pi_{;\mu\nu}\pi^{;\mu\nu} \right) -\frac12 \pi_{;\mu}\pi^{;\mu} R \right) \right], \nonumber
\ee
 where $\bar{\kappa} \equiv c_4 C^2/r_g^2$.
In this case the problem of accretion becomes quite cumbersome. 
To start with, let us first consider the case when only $\L_4$ term is present in the action (\ref{action0}).
%
\subsection{No canonical term, $\e=0$.}
\begin{figure}[t]
\begin{center}
\includegraphics[width=0.5\textwidth]{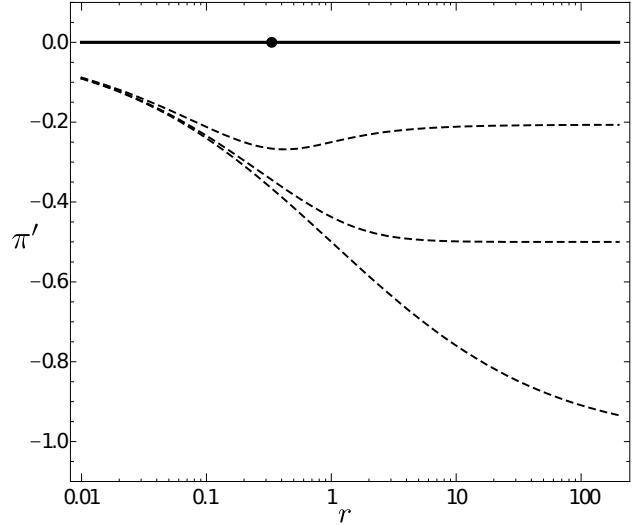}
\caption{Case, $\L\sim  \L_4$. Solutions for $\pi' = -1/f+\psi'(r)$ are shown for various values of $A$.
The critical flow is for $A=8$ (solid thick line), the critical point is indicated by the black dot, $r_* = 1/3$.
For the cases $A= 4$, $1$, $0$ the solution is shown by dotted lines, from top to bottom.
}
\label{fig L4 A}
\end{center}
\end{figure}

%
One can use the fact that the energy-momentum tensor for $\L_4$ is conserved, $T^{\mu\nu}_{\hphantom{\mu;};\nu}=0$ \cite{Deffayet:2009wt}.
Since $T^\mu_\nu$ does not depend on $v$ explicitly for the ansatz (\ref{ansatz}), we obtain,
\be
\label{consT}
T_v^r = \frac{A}{r^2},
\ee
where from (\ref{emt4}), applying the ansatz (\ref{ansatz}), after lengthy but straightforward calculations we find,
\be
\label{emt4ans}
T^{r}_v = -\frac{2\psi'}{r^2} \left(-r^2 f^2 f'' \psi'^2+r^2 f''-6 r f' \left(f^2 \psi'^2-1\right)-4 f^3 \psi'^2\right),
\ee
where without loss of generality we have set the coefficient in front of $\L_4$ to unity.
Now, the problem of accretion for non-zero $\L_4$ (and $\L_i=0$ for $i\neq 4$) is reduced to the solving (\ref{consT}) and (\ref{emt4}) 
and choosing the physical solution, if any.
Combining (\ref{consT}) and (\ref{emt4ans}) we obtain three branches of solutions $\psi'_{(4)}$. 
For $A<0$ or $A>8$ there is no well-behaved solution.
It is interesting that for any $A$ in the range $0 \leq A  <8$ a smooth solution exists for $\pi'$, covering the whole space, $0<r<\infty$, with 
corresponding characteristics finite everywhere, see Fig.~\ref{fig L4 A}. The critical point is absent and the solution at infinity is,
\be
\label{L4 sol inf}
\psi'_{(4)} = \frac{A^{1/3}}{2} + \frac{2+ A^{2/3}}{3 A^{1/3} r} + {\cal O}\left(\frac{1}{r^2}\right), \quad 0 \leq A  <8. \nonumber
\ee
The critical point exists for $A=8$ at $r_* = 1/3$; one can check that the sound horizon is indeed at $r=1/3$. 
The solution critical solution is then,
\be
\label{L4 crit sol inf}
\psi'_{(4),*} = \frac{r}{r-1}, \nonumber
\ee
which implies $\pi'_{(4),*} = 0 $, see Fig.~\ref{fig L4 A}.

\begin{figure}[t]
\begin{center}
\includegraphics[width=0.5\textwidth]{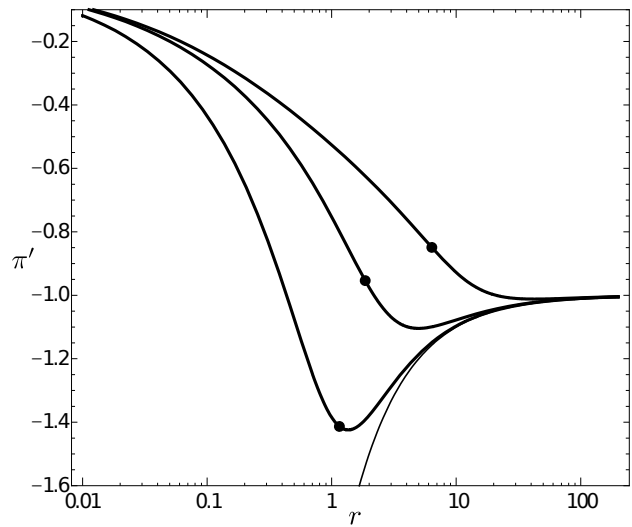}
\caption{Case, $\L\sim \L_2+ \L_4$. Solutions for $\pi' = -1/f+\psi'(r)$ are shown for different $\bar{\kappa}$:
 $\bar{\kappa}=-100$, $-1$,  $-0.04$ (thick lines, from the top to bottom) and 
 $\bar\kappa=0$, i.e.  only canonical term (thin line).
The positions of sound horizon are shown by the dots,
for $\bar\kappa <0$ they are outside the Schwarzschild horizon. For $\bar\kappa <0$   solutions are singular.}
\label{fig L4}
\end{center}
\end{figure}
%
\subsection{Generic case, $\e\neq 0$, $\bar{\kappa}\neq 0$.}
%
Let us now consider the case when besides the term $\L_4$ in the action, there is also the canonical kinetic term, $\L\sim \L_2 + \L_4$.
Depending on the signs in front of $\L_2$ and $\L_4$, one can distinguish, in principle, four different cases. 
Let us first take $\L_2$ to be canonical, i.e. $c_2=1$ and also $c_4>0$. 
In this case a family of smooth solutions exists for a certain range of positive $A$, $0<A\leq A_*(\bar\kappa)$. 
For negative $A$ the solutions diverge at the Schwarzschild horizon.
The critical solution, realized for $A=A_*(\bar\kappa)$, is regular for $0<r<\infty$ with the critical surface inside the Schwarzschild horizon.
Interestingly, however, that even though smooth background solutions  exist  for $0<A\leq A_*(\bar\kappa)$, 
the propagation vectors are nonexistent for some radii outside the Schwarzschild horizon. 
In fact, the equation of motion for perturbations become elliptic, 
indicating that solutions are catastrophically unstable.

On the other hand, for $\e=1$ and $\bar{\kappa}<0$, a critical solution exists,
with well-behaved propagation vectors, at least up to the sound horizon. 
The asymptotic form of the physical solution is,
\be
\label{can L4 sol inf}
\psi'_{(2,4)*} = \frac{A}{2r^2} +  {\cal O}\left(\frac{1}{r^3}\right). \nonumber
\ee
Note that the leading asymptotic behavior at infinity is the same as in the case of canonical term, up to the difference in constant $A$, which is simply 
the reflection of the fact that at infinity the term $\L_4$ becomes unimportant compared to $\L_2$.

\section{Results and discussion}

The results of the paper can be summarized as follows. 
For the Lagrangian (\ref{L2L3action}), $\L \sim \L_2 + \L_3$, we have found,
	\begin{itemize}
	\item
	case $c_2=0$  (canonical kinetic term is absent). Critical solution exists, Eq.~(\ref{solNCph}), with the critical point at $r_*=3/4$, 
	see Fig.~\ref{fig DGP}, upper (red) line.
	\item 
	case $c_3/c_2>0$. Critical solution exists, with asymptotic form (\ref{sol2inf}) at infinity, the critical radius $r_*$ is inside  
	the Schwarzschild horizon.
	\item
	case $c_3/c_2<0$. Critical solution exists as well, with the asymptotic form (\ref{sol2inf}) at infinity, however the critical radius $r_*$ is ouside  
	the Schwarzschild horizon.
	\end{itemize}

For the Lagrangian $\L \sim \L_2 + \L_4$,
	\begin{itemize}
	\item
	case $c_2= 0$ (no kinetic term). There is a family of solutions parametrized by the value total influx, $A$. 
	These solutions and the corresponding propagation vectors are smooth for $0<r<\infty$.
	Critical solution also exists for a particular value of the influx (for $A=8$), with the critical point at $r_* = 1/3$, see Fig.~\ref{fig L4 A}
	\item
	case $c_4/c_2>0$. A family of smooth solution exist for some range of positive $A$. The critical solution also exists with a critical point $r_*<1$. 
	However, for the found solutions the propagation vectors do not exist in some region of $r$, $r>1$.
	\item
	case $c_4/c_2<0$. There is the well-behaved critical solution and propagation vectors, corresponding to the solution. 
	The sound horizon is outside the Schwarzschild horizon, see Fig.~\ref{fig L4}.
	\end{itemize}

It is interesting to see if we get any new constraints on the Lagrangian compared to 
those obtained for the non-covariant version of galileon.
Since in our study the space-time is asymptotically flat, it would be natural to apply constraints on $c_i$ obtained in \cite{Nicolis:2008in} for 
flat space-time\footnote{Then in \cite{Nicolis:2008in} one should make substitution $d_i\to c_i$.}. 
For the Lagrangian containing only $\L_2$ and $\L_3$ we did not find any constraints on the coefficients $c_2$, $c_3$.
Note that $c_2>0$ was found in \cite{Nicolis:2008in} by requiring the model to be ghost-free.
Since we did not address the issue of ghosts in our study\footnote{In fact, the presence of ghost does not affect the solutions we found, since the overall sign in front of the action is unimportant.}, it is not surprising that we did not find any constraints on $c_2$.
It is more interesting, that if we impose positivity of $c_2$ we still do not get any constraints on $c_3$ 
(meantime the study in \cite{Nicolis:2008in} implies that $c_3>0$ if applied to our case).
There is a subtle point here: for our solution the case $c_3/c_2<0$ corresponds to the critical point outside the Schwarzschild horizon,
and formally the solution diverges at some point inside both horizons. 
This behavior, however, cannot be used to invalidate the solution, because the region inside of both horizons is causally disconnected
from the asymptotically free region, 
so, in principle, inside the Schwarzschild horizon the solution can change to another one (the one which does not satisfy the steady-state ansatz).

For the Lagrangian $\L\sim \L_2 + \L_4$ we also obtained an interesting result. 
In particular, for both $c_2$ and $c_4$ positive (and $c_3=0$),
the solution for the steady-state accretion is unstable, because the propagation vectors do not exist in some interval of $r$ outside the Schwarzschild 
horizon. On the other hand, for $c_4/c_2 <0$, we obtained a well-behaved critical solution with corresponding propagating vectors.
These results are in contrast to those obtained in \cite{Nicolis:2008in} for a static configuration of  non-covariant galileon, where
for $c_3 =0 $ and asymptotically flat space-time the coefficient $c_4$ has to be positive or zero.

The propagation of small perturbations for galileon is superluminal for a range of parameters. 
Thus, it is not surprising that in this range of parameters the sound horizon is inside the Schwarzschild horizon,
implying that one can in principle communicate with the interior part of a black hole, $r>r_*$. 
The same effect was found for ``superluminal'' k-essence \cite{Babichev:2006vx,Babichev:2007wg}.
Similarly to k-essence accretion, the acoustic space-time for accreting galileon is globally hyperbolic, in spite of the superluminal perturbations. 
This can be explicitly demonstrated by constructing a Schwarzschild-like metric of the perturbations of the field. 
We take the case $\L \propto \L_3$ for simplicity\footnote{A similar arguments can be repeated for other models of galileon, considered above.}.
Substituting (\ref{EF}) and (\ref{solNCph}) into (\ref{G}) one can find the effective contravariant metric. By inverting the obtained metric,
one finds the covariant effective metric $\tilde{G}_{\m\n}$.
Making  successive change of coordinates,  $dv = dT - G_{tr}/G_{tt} dr$, $r = R^{3/4}$ and $T\to \sqrt{12} T$, we finally get,
\be
\label{eff S}
dS^2 = - (4R^{4/7} - 3) R^{2/7} dT^2 + \frac{R^{4/7}}{4R^{4/7} -3} dR^2 + R^2d\Omega.
\ee
The above expression for the effective metric is similar to the Schwarzschild metric, with the important difference that the 
position of the event horizon is now at $R^{4/7}= 3/4$.

Although the global hyperbolicity is not affected by the presence of the superluminal propagation in this case of steady-state 
accretion of galileon, it is still unclear, however, 
how to embed the existence of  the new ``event'' horizon into the standard thermodynamics of black holes.
In particular, since the sound horizon is inside the Schwarzschild one, the latter loses its meaning of the universal event horizon, 
therefore one may need to modify the entropy assigned to a black hole.

It is interesting to note, that for a range of parameters, our solutions for accretion possesses the analogue of the Vainshtein behavior, similar to the case of 
static sources.\footnote{Note, however, that in order to observe the Vainshtein effect, 
the matter must be coupled to the galileon non-minimally, so that the fifth force (due to the direct coupling) leads to the deviation from 
General relativity on large scales, while on small scales the galileon is screened. 
Meanwhile, our results for the accretion do not depend on coupling to matter, since the solution is matter-free. } 
Indeed, let us take the case $\L\sim \L_2 + \L_3$ for definiteness and $\e>0$. In the limit $r\to \infty$ the solution is given by (\ref{sol2inf}),
which has the same form as the solution for the standard kinetic term (\ref{solst}), apart from the overall constant $A\to A+\kappa$.
In this regime the kinetic term is dominant, and assuming that non-minimal coupling is given by a term $\sim \pi T/M_P$ 
(in the Einstein frame) a test particle would experience an additional force, $F_\pi$, which is to be compared to the 
gravitational force $F_{\rm grav}$,
\be
\label{5th}
\frac{F_\pi}{F_{\rm grav}} \sim \frac{\nabla\pi}{M_P}\frac{r^2}{r_g} \sim \frac{\dot\pi_\infty r_g (A+\kappa)}{M_P},
\ee
where we restored the physical units. 
On the other hand, if also $\kappa>0$, then the solution changes its behavior at the radius,
\be
\label{Vainshtein}
r_\star\simeq  r_g\left[ 4\kappa (A+\kappa)\right]^{1/3}.\nonumber
\ee
where $r_\star$ is a analogue of the Vainshtein radius. The ``accretion'' Vainshtein radius is related to the 
``usual'' one, $r_V$, defined for a static source as follows,
\be
\label{VV}
\frac{r_\star}{r_V} \sim \left(\frac{\dot\pi_\infty r_g}{M_P}\right)^{1/3}.\nonumber
\ee
Thus, for the range of distances $r_g < r < r_\star$, the ratio of the galileon and gravitational forces is,
\be
\label{5thscreened}
\frac{F_\pi}{F_{\rm grav}} \sim  \frac{\dot\pi_\infty r_g (A+\kappa)}{M_P} \left(\frac{r}{r_\star}\right)^{3/2},\nonumber
\ee
which has an additional small factor $(r/r_\star)^{3/2}$ compared to (\ref{5th}), being a manifestation of the Vainshtein effect.
This can also be seen in Fig.~(\ref{fig DGP}): the behavior of $\pi'$ for $\kappa>0$ is smoother closer to the horizon, 
compared to the canonical kinetic term. 
Note, that if  $\e$ and $\kappa$ have different signs (e.g. $\e>0$ and $\kappa<0$), then 
solutions do not exhibit  the Vainshtein behavior, due to the fact that the term in (\ref{sol12}), 
which responsible for this (the second term inside the square root) is negative and therefore it can never 
be dominant.
A similar analysis can be made for the Lagrangians of the form $\L\sim \L_2 + \L_4$, and one can find that 
the solutions have the Vainshtein behavior in this case as well.

It is worth to mention the change of the black hole mass during the accretion of the galileon. 
Since by definition the problem we studied was steady-state in the test-fluid approximation, 
the rate of the change can be found via the total flux at the infinity, $r\to \infty$.
In the Schwarzschild coordinate system the total flux is $\propto (r^2 T_t^r)$. 
Expressing $T_t^r$ through the components in the EF coordinates, 
we obtain\footnote{The same expression can be obtained by calculating 
the flux at of the Killing energy across the Schwarzschild horizon, $dM/dt = Area\times T_{vv}$.}, 
\be
\label{BHmass}
\frac{dM}{4\pi dt} =  A\, r_g^2 \dot{\pi}^2_\infty ,
\ee
where the physical units have been restored.
In particular, for the case when only the canonical term is present, $A=2$, one gets the known result (compare, i.e. with \cite{Babichev:2004yx,Babichev:2005py},
the additional factor 2 in the above expression is due to the choice of the canonical term $(\pd\pi)^2$, instead of $1/2(\pd\pi)^2$). 
The flux can be easily made negative by changing of the overall sign of the Lagrangian for $\pi$, so that the black hole will loose mass instead of 
gaining it. Usually, such a change of sign is accompanied by ghosts. 
However, e.g. in \cite{Deffayet:2010qz} it was shown that even though the $\L_2$ term is ghost-like, 
nevertheless the full Lagrangian, $\L\sim\L_2 + \L_3$, is ghost-free
in the vicinity of the cosmological attractor.
One can think of exploiting this particular feature to construct negative flow onto a black hole. 
In this case, however, one immediately stumbles the following problem. 
In our study the space-time is taken to be asymptotically flat, implying that the term $\L_3$ is subdominant 
in comparison to $\L_2$, implying the  appearance of ghosts in the model studied in \cite{Deffayet:2010qz}.
(Note that the expansion of the universe in the attractor is the key to to avoid ghosts in \cite{Deffayet:2010qz}). 
Thus, in the region of space-time where the gravitation of a black hole is stronger than  the cosmological expansion, 
one may naively expect the appearance of ghost.
This question, however, deserves a separate and more careful study.

Let us in conclusion discuss the assumptions we made to find the above results. 
First of all we assumed that the accretion is steady-state, which is physically reasonable when the Lagrangian is shift-symmetric and 
it is enough time for the system to evolve before the steady-state is established.
We also neglected the backreaction of the accreting fluid on the metric, since
the backreaction can be made parametrically small. 
Indeed, the total flux onto the black hole is proportional to $r_g^2 \dot{\pi}_\infty^2$ (\ref{BHmass}). 
In the limit $c_3 = {\rm const}$, $r_g\to 0$, $\dot{\pi}_\infty\to 0$ (and keeping $\dot{\pi}_\infty/r_g = {\rm const}$, so that the 
solution, expressed  in dimensionless units is not affected\footnote{Here we restored the physical units, so that $\dot{\pi}$ has dimension 2.}), 
the total flux goes as $\sim r_g^4$, while the mass of the black hole $\sim r_g$, 
thus validating our the assumption. 
Another source of possible inaccuracy in our calculations is that we also neglected the kinetic mixing term  of perturbations and gravity.
E.g. the term $\sim\pd\pi\pd\pi\delta R$ was neglected in (\ref{G}). The contribution of this term is to compared to 
the contributions of the terms left over in (\ref{G}), one can check that in order to neglect the backreaction of the perturbations,
one needs $(\pd\pd\pi) \gg c_3 \left(\pd\pi\right)^4/M_P^2$, which again can be maid parametrically small by taking the limit we used before.

To summarize, in this paper we studied the steady-state accretion of the covariant galileon onto a Schwarzschild black hole.
Subject to the assumptions we made, our study of existence and classical stability of solutions did not show 
any constraints on the parameters of the Lagrangian of the form $\L\sim \L_2 + \L_3$. 
In the event when $ \L_4$ is present,
the constraints on the parameters of the theory we obtained are in a sense orthogonal to the constraints found in other studies.
We also notice that for some range of parameters of the theory signals can be sent from the inside of the Schwarzschild horizon, 
being a consequence of superluminal propagation of perturbations.
The accretion of the energy flow can be made negative by appropriate choice of the coefficients in the Lagrangian.

\section*{Acknowledgments}
I would like to thank Slava Mukhanov for discussions and Alex Vikman for  discussions, interesting suggestions and a critical reading of the manuscript.
I would also like to thank the referee for a suggestion to check the Vainshtein behavior of the solutions. 
The work was supported by the TRR 33 ``The Dark Universe''.

\bibliographystyle{apsrev4-1}
\bibliography{bibliography}

\end{document}